\documentclass[a4paper,11pt]{article}
\pdfoutput=1 

\usepackage{jcappub} 
\usepackage[normalem]{ulem} 

\usepackage{amsmath}
\usepackage{amsfonts,color}
\usepackage{amssymb,float}
\usepackage{mathtools}
\usepackage[utf8]{inputenc}
\usepackage{graphicx}
\usepackage{wasysym}
\usepackage{tabularx}
\usepackage{accents}
\usepackage{color, colortbl}
\usepackage[dvipsnames]{xcolor}
\usepackage{float}

\title{\boldmath Implications of pulsar timing arrays \\ for Gauss-Bonnet Inflation}

\author[a]{Reginald Christian Bernardo,}
\author[b]{Seoktae Koh,}
\author[b]{\\ and Gansukh Tumurtushaa~\footnote{All authors contributed equally to this work.}}

\affiliation[a]{Asia Pacific Center for Theoretical Physics, Pohang 37673, Korea}
\affiliation[b]{Department of Science Education, Jeju National University, Jeju, 63243, Korea}

\emailAdd{reginald.bernardo@apctp.org}
\emailAdd{kundol.koh@jejunu.ac.kr}
\emailAdd{gansukh@jejunu.ac.kr}

\abstract{
Correlated time-of-arrival measurements by pulsar timing arrays (PTAs) have provided a new means of constraining astrophysical or cosmological models that produce a gravitational wave (GW) background. For this work, we discuss the implications of PTA observations for Gauss-Bonnet (GB) inflationary models through the production and propagation of inflationary GWs. We show that our GB inflationary scenario is consistent with present PTA and cosmological data. A blue-tilted tensor power spectrum supported by PTAs can be naturally accommodated in GB inflation. Using observational constraints, we derive general conditions for the inflaton potential and the GB coupling function, suggesting that in GB inflation, the inflaton must climb up the potential before rolling downhill and reaching the end of inflation. We provide two concrete GB inflationary models to demonstrate the viability of this mechanism.
}

\begin{document}
\maketitle

\section{Introduction}\label{sec:1}
Compelling evidence for a nanohertz stochastic gravitational wave background (SGWB) from pulsar timing arrays (PTAs) \cite{NANOGrav:2023gor, Reardon:2023gzh, EPTA:2023fyk, Xu:2023wog, Miles:2024seg} has marked a major milestone in gravitational wave (GW) astronomy, astrophysics, and cosmology, following the first direct detections of GWs from compact binaries by ground-based detectors \cite{LIGOScientific:2016aoc, KAGRA:2021vkt, LIGOScientific:2021sio}. A crucial aspect of the nanohertz SGWB is its origin, which arises from the superposition of GWs from numerous unresolved sources \cite{Joshi:2013at, McLaughlin:2014wna, Manchester:2015mda, Lommen:2015gbz, Romano:2016dpx, Becker:2017yyc, Verbiest:2021kmt, Taylor:2021yjx, Verbiest:2024nid, Yunes:2024lzm, Bernardo:2024bdc, Domcke:2024soc}. The prevailing interpretation attributes the signal to an ensemble of inspiralling supermassive black hole binaries (SMBHBs) \cite{Sazhin:1978myk, Detweiler:1979wn, Phinney:2001di, Wyithe:2002ep, Sesana:2004sp, Sesana:2008mz, Vigeland:2016nmm, Burke-Spolaor:2018bvk, Liu:2021ytq, Sato-Polito:2023gym, Ellis:2023dgf, Sato-Polito:2023spo, Bi:2023tib, Sato-Polito:2024lew, Sah:2024oyg, Raidal:2024odr, Sah:2024etc, Ellis:2024wdh}. However, the current data leave room for a variety of alternative explanations, including scenarios involving the early universe and physics beyond the standard model \cite{Chen:2019xse, Ellis:2020ena, Vagnozzi:2020gtf, Samanta:2020cdk, Benetti:2021uea, NANOGrav:2021flc, Buchmuller:2021mbb, Deng:2021gkx, Domenech:2021ztg, Bari:2021xvf, Xue:2021gyq, Sharma:2021rot, Hindmarsh:2022awe, EPTA:2023xxk, Vagnozzi:2023lwo, Figueroa:2023zhu, Ellis:2023dgf, Saeedzadeh:2023biq, Liu:2023pau, Liu:2023hpw, Chen:2023bms, Liu:2023ymk, Jin:2023wri, Gangopadhyay:2023qjr, Huang:2023chx, Ye:2023tpz, Wang:2023sij, Wang:2023ost, Zhu:2023lbf, Jiang:2023gfe, Bian:2023dnv, Bari:2023rcw, Datta:2023xpr, Datta:2023vbs, Jiang:2024dxj, Winkler:2024olr, Agazie:2024kdi, Calza:2024qxn, Papanikolaou:2024fzf, Papanikolaou:2024cwr, Kume:2024adn, Basilakos:2024diz, Chen:2024fir, Chen:2024twp, Datta:2024bqp, Athron:2024fcj, Domenech:2025bvr, Yogesh:2025hll}. This has established nanohertz GWs as a novel probe of previously inaccessible energy scales, offering the first constraints on primordial GW sources that can contribute to the nanohertz frequency band.

Among these alternative explanations, inflation \cite{Guth:1980zm, Starobinsky:1980te, Sato:1980yn, Linde:1981mu, Albrecht:1982wi}---a period of accelerated expansion in the early Universe---arguably stands out as particularly compelling. Originally proposed to explain the observed homogeneity of the cosmos and the origin of primordial density perturbations consistent with the CMB \cite{WMAP:2008lyn, WMAP:2010qai, WMAP:2012nax, Planck:2013jfk, Planck:2015sxf, Planck:2018jri, Planck:2018vyg,  Planck:2019evm}---inflation naturally produces an SGWB spanning a wide range of frequencies, including the nanohertz band probed by PTAs \cite{Starobinsky:1979ty, Allen:1987bk, Sahni:1990tx}. Despite strong support from traditional astronomical observations, definitive evidence for inflation, a so-called `holy grail,' remains elusive, in the form of CMB B-modes, or primordial GWs in the CMB polarization \cite{Kamionkowski:2015yta, LiteBIRD:2023zmo, SimonsObservatory:2025wwn}. In this context, PTA data offers a complementary avenue to test inflation: even if the observed nanohertz SGWB originates from an astrophysical source, PTA measurements can still provide meaningful upper bounds, guiding future theoretical developments. In fact, current PTA observations have already placed constraints on inflationary phenomena, see e.g., \cite{Vagnozzi:2020gtf, Sharma:2021rot, NANOGrav:2023hvm, EPTA:2023xxk, Figueroa:2023zhu, Jiang:2023gfe, Ye:2023tpz, Niu:2023bsr, Vagnozzi:2023lwo, Bian:2023dnv, Ellis:2023dgf, Tan:2024kuk}, allowing for rigorous tests of its most fundamental prediction of a nearly Gaussian, scale-invariant spectrum of primordial perturbations.

Beyond its cosmological significance, inflation also provides a window into early universe gravitational physics, as modifications to the inflationary dynamics can alter the production of primordial tensor perturbations. In the standard inflationary framework based on general relativity (GR), a single scalar field with a nearly flat potential drives a quasi-de Sitter expansion, giving rise to a slightly red-tilted power spectrum for primordial tensor modes. This is due to the so-called consistency relation $n_T=-r/8$, where $n_T$ denotes the spectral tilt of the tensor power spectrum while $r$ is known as the tensor-to-scalar ratio. The upper bound on the tensor-to-scalar ratio has been given as $r<0.036$ (at the $95\%$ confidence level)~\cite{BICEP:2021xfz}. The tensor spectral index with $n_T=0$ corresponds to a scale-invariant spectrum, and $n_T<0$ ($n_T>0$) corresponds to a red (blue) spectrum. Thus, a blue-tilted spectrum suggests deviations from standard inflation, hinting at modifications to gravity at inflationary energy scales. Interestingly, current PTA data appears to favor a blue-tilted tensor power spectrum, motivating a closer examination of inflationary models that naturally predict such a feature. This can be accommodated in string gas cosmology~\cite{Brandenberger:2006xi}, super-inflation models~\cite{Baldi:2005gk}, non-commutative inflation~\cite{Calcagni:2004as, Calcagni:2013lya}, and inflationary particle production~\cite{Cook:2011hg, Mukohyama:2014gba}. In this work, we explore the implications of PTA data for Gauss-Bonnet (GB) inflation, a well-motivated extension of GR that introduces a coupling between the inflaton and the GB term (Section \ref{subsec:background_dynamics}).

Non-standard inflation with a modified gravitational sector can often be captured within the G-inflation framework \cite{Kobayashi:2011nu, Deffayet:2009wt, Deffayet:2011gz} or, more generally, Horndeski theory \cite{Horndeski:1974wa, Kobayashi:2019hrl}. While kinetically driven models of G-inflation \cite{Kobayashi:2010cm} can produce a blue-tilted GW spectrum, potential-driven models more commonly predict a red tilt \cite{Kamada:2012se}. Although exceptions exist, a red-tilted spectrum is often considered the default expectation. GB inflation stands out as a notable counterexample, as it naturally generates a blue-tilted GW spectrum \cite{Satoh:2008ck, Satoh:2010ep, Guo:2010jr, Jiang:2013gza, Koh:2014bka, vandeBruck:2016xvt, Bhattacharjee:2016ohe, Koh:2016abf, Nozari:2017rta, Wu:2017joj, Nojiri:2017ncd, Koh:2018qcy, Chakraborty:2018scm, Tumurtushaa:2018lnv, Mishima:2019vlh, Koh:2023zgn, Odintsov:2023weg}. Given the recent and future PTA data, there is a motivation to examine the viability of GB inflation \cite{Odintsov:2020xji, Odintsov:2020sqy, Odintsov:2023lbb, Nojiri:2023mvi, Odintsov:2023aaw} through primordial GWs \cite{Yin:2024ccm, Chen:2025wcw, Gao:2025slp}.

First results were presented in \cite{Yin:2024ccm} that constrained the steepness of the potential and the GB coupling, both power laws, by using PTA data; in particular, it was shown that the potential was constrained to be of the inverse power law form, contrary to usual assumptions. However, in this work, we take a different route, using a phenomenological GW spectrum and determining necessary conditions for the potential and the coupling to satisfy PTA and cosmological data. 

The rest of this work is organized as follows. We start with a brief overview of standard inflation and PTA constraints (Section \ref{sec: 2}). Then, through this lens, we discuss GB inflation, particularly how we deal with the generation and propagation of GWs. We present constraints on its parameter space with PTA and cosmological data (Section \ref{sec: 3}). Lastly, we provide two concrete GB models that partly satisfy observational constraints or newly determined general conditions based on observational constraints (Section \ref{sec: 4}). Throughout this work, we use the mostly-plus metric signature $(-,+,+,+)$ and geometrized units $G=c=1$.

\section{Inflationary Gravitational Waves and Pulsar Timing Arrays} \label{sec: 2}

We give a brief pedagogical overview of GWs produced during inflation on scales relevant to PTA experiments~\cite{Kuroyanagi:2014nba}. The expert/inflationary theorist may skip Section \ref{subsec:inflationary_gws_at_nanohertz} and jump to Section \ref{subsec:pta_cosntraints_standard_inflation} on PTA constraints on standard inflation.

\subsection{Inflationary GWs at nanohertz regime}
\label{subsec:inflationary_gws_at_nanohertz}

Let us start with a tensor part of the metric perturbation in the flat Friedmann-Lemaitre-Robertson-Walker (FLRW) metric, whose perturbed line element in synchronous gauge is given by
\begin{align}\label{eq: metricT}
    ds^2=a^2(\tau)\left[-d\tau^2 +(\delta_{ij} +h_{ij}(\tau,\textbf{x}))dx^idx^j\right]\,,
\end{align}
where $a$ and $\tau$ denote the scale factor and conformal time, respectively. GWs correspond to the transverse ($h_{ij,j}=0$) and traceless ($h^{i}_i=0$) part of the metric perturbation $h_{ij}$.  The average GW energy density $\bar{\rho}_{\rm GW}$ is given by 
\begin{align}
    \bar{\rho}_{\rm GW} = \frac{M_{\rm pl}^2}{8a^2}\langle (h_{ij}')^2+(\nabla h_{ij})^2\rangle\,,
\end{align}
where $M_{\rm pl}$ is the reduced Planck mass, and $\langle \dots\rangle$ indicates an ensemble average. By Fourier transforming $h_{ij}(\tau, \textbf{x})$ as 
\begin{align}\label{eq: fourierTrans}
    h_{ij}(\tau, \textbf{x}) = \sum_{\lambda=+, \times}\int \frac{dk^3}{(2\pi)^{3/2}}\epsilon_{ij}^\lambda h_{\textbf{k}}(\tau) e^{i\textbf{k}\cdot\textbf{x}}\,,
\end{align}
where the polarization tensor $\epsilon_{ij}^\lambda$ satisfies the transverse-traceless (TT) conditions and is normalized by $\sum_{i,j} \epsilon^\lambda_{ij} (\epsilon_{ij}^{\lambda'})^\ast=2\delta^{\lambda \lambda'}$, one can rewrite
\begin{align}\label{eq: GWrho}
    \bar{\rho}_{\rm GW} = \frac{M_{\rm pl}^2}{4}\int d\ln k \left(\frac{k}{a}\right)^2\frac{k^3}{\pi^2}\sum_\lambda\left| h_{\textbf{k}}^\lambda\right|^2\,.
\end{align}
An important quantity relevant for GW detection is the energy density parameter $\Omega_{\rm GW}$, i.e., the ratio between the energy density of GWs per logarithmic interval of wavenumber (or frequency) today and the critical density $\rho_{\text{crit}}=3H^2M_{\rm pl}^2$; in symbols, this is given by
\begin{align}
    \Omega_{\rm GW}(k) = \frac{1}{\rho_\text{crit}}\frac{d\bar{\rho}_{\rm GW}}{d\ln k}\,.
\end{align}
The value of the Hubble parameter today is measured as $H_0=100 h_0\, \text{km}\, \text{s}^{-1}\, \text{Mpc}^{-1}$ with $h_0=0.674 \pm 0.005$ \cite{Planck:2018vyg}. With \eqref{eq: GWrho}, we can write
\begin{align}\label{eq: OmGW}
    \Omega_{\rm GW}(k) = \frac{k^2}{12a^2H^2}P_T(k)\,,
\end{align}
where the present GW power spectrum $P_T(k)$ is related to its inflationary counterpart, the primordial tensor power spectrum, $\mathcal{P}_T(k)$, via a transfer function $\mathcal{T}(k)$, effectively encoding cosmological processes that underwent between the end of inflation and today, i.e., the present GW power spectrum is given by
\begin{align}\label{eq: Pt}
    P_T(k) \equiv  \frac{k^3}{\pi^2}\sum_\lambda \left| h^{\lambda}_\textbf{k}\right|^2=\mathcal{T}^2(k)\mathcal{P}_T(k)\,.
\end{align}
The important pieces that chip in to the observable, $\Omega_{\rm GW}(f)$, are the primordial tensor power spectrum and the transfer function. In the following, we briefly describe how both are derived. Firstly, the primordial tensor power spectrum is usually considered as a power law,  
\begin{align}\label{eq: infPt}
    \mathcal{P}_T(k) = \mathcal{P}_T(k_\ast) \left(\frac{k}{k_\ast} \right)^{n_T}\,,
\end{align}
where $\mathcal{P}_T(k_\ast)$ and $n_T$ are the amplitude and the spectral tilt at the reference scale $k_\ast$. The amplitude $\mathcal{P}_T(k_\ast)$ is often quantified in terms of the tensor-to-scalar ratio $r$ as $\mathcal{P}_T(k_\ast)=r\, \mathcal{P}_S(k_\ast)$, where the amplitude of the scalar power spectrum $\mathcal{P}_S(k)$ of scalar perturbations is well measured by observation as $\ln \left(10^{10}\mathcal{P}_S \right)=3.089^{+0.024}_{-0.027}$ at $k_\ast=0.05\,\text{Mpc}^{-1}$ \cite{Planck:2018jri, Planck:2018vyg}.~\footnote{More general parametrizations of the power spectrum have been used~\cite{Planck:2018jri, Planck:2018vyg}, but we shall be content with the power law form for this work.} Since the actual spectrum of primordial GW is unknown, one can consider modified spectrum forms, such as broken power-law form, as discussed in \cite{Jiang:2023gfe, Benetti:2021uea}. For simplicity, we choose a power-law form as presented above.

The second important piece is the transfer function, reflecting the evolution of GWs after the relevant GW modes cross the horizon. Thus, it depends on the thermal history of the Universe, and its form can be obtained by numerically solving the perturbed equations of motion for GWs, or rather in GR,
\begin{align}\label{eq: hGR}
    h_{k}''+2\mathcal{H}h'_{k}+k^2h_{k} =0\,,
\end{align}
where $\mathcal{H}\equiv a'/a$ with the prime denoting the derivative with respect to conformal time $\tau$ and $k$ is the wavenumber. Depending on when the mode crosses the horizon, the solutions to \eqref{eq: hGR} have qualitative behavior in two regimes either outside the horizon ($k\ll \mathcal{H}$) where the amplitude of $h_{k}$ remains constant or inside the horizon ($k\gg\mathcal{H}$) where the amplitude begins to damp due to cosmic expansion. 

A widely acceptable post-inflationary scenario consists of matter domination (MD)-like phase before reheating, followed by radiation domination (RD)~\cite{Kuroyanagi:2014nba}.
In this case, the transfer function can be broken down as follows
\begin{align}\label{eq: transF}
    \mathcal{T}^2(k) = \Omega_m^2\left(\frac{g_\ast(T_\text{in})}{g_{\ast 0}}\right)\left(\frac{g_{\ast s0}}{g_{\ast s}(T_\text{in})}\right)^{4/3}\left(\frac{j_1(k\tau_0)}{k\tau_0}\right)^2\mathcal{T}^2_{1}\left(\frac{k}{k_\text{eq}}\right) \mathcal{T}^2_{2}\left(\frac{k}{k_\text{rh}}\right)\,,
\end{align}
where $\Omega_{m}$ is the matter density of the universe, $g_\ast(T_\text{in})$ and $g_{\ast s}(T_\text{in})$ are the effective number of relativistic degrees of freedom and their counterpart for entropy, respectively, and $T_\text{in}$ is the temperature of the universe when the mode $k$ re-enters the horizon. A subscript $0$ denotes present-day quantities. The scale dependence of $g_{\ast}$ is approximately expressed as~\cite{Kuroyanagi:2014nba}~\footnote{{To incorporate the temperature dependence of the relativistic degrees of freedom in the GW spectrum~\cite{Watanabe:2006qe}, Ref.~\cite{Kuroyanagi:2014nba} introduced the fitting function given in Eq.~(\ref{eq: gstarss}).}}
\begin{align}\label{eq: gstarss}
    g_{\ast}(T_\text{in}(k))= g_{\ast 0} \left(\frac{A+\tanh k_1}{1+A}\right)\left(\frac{B+\tanh k_2}{1+B} \right)\,,
\end{align}
where
\begin{align*}
    A = \frac{-1-10.75/g_{\ast 0}}{-1+10.75/g_{\ast 0}}\,,\quad 
    B = \frac{-1-g_\text{max}/10.75}{-1+g_\text{max}/10.75}\,,
\end{align*}
and
\begin{align*}
    k_1 = -2.5 \log_{10}\left(\frac{k/2\pi}{2.5\times 10^{-12}\,\text{Hz}}\right)\,,\quad k_2 = -2.0 \log_{10}\left(\frac{k/2\pi}{6.0\times 10^{-9}\,\text{Hz}}\right)\,,
\end{align*}
with $g_\text{max}=106.75$. The same formula can be applied for the counterpart for entropy $g_{\ast s}(T_\text{in}(k))$ by replacing $g_{\ast 0 }=3.36$ with $g_{\ast s0}=3.91$. 
The wavenumbers corresponding to matter-radiation equality and the completion of reheating are, respectively, 
\begin{align*}
    k_\text{eq}=7.3 \times 10^{-2}\Omega_m h_0^2\,\text{Mpc}^{-1}\,,\quad {\rm and} \quad
    k_\text{rh}=1.7 \times 10^{13}\left(\frac{g_{\ast s}(T_\text{rh})}{106.75}\right)^{1/6}\left(\frac{T_\text{rh}}{10^6\text{GeV}} \right)\,\text{Mpc}^{-1}\,.
\end{align*}
Modes that enter the horizon during the late-time MD phase evolve as $h_{k}\sim 3j_1(k\tau_0)/(k\tau_0)$, where $j_l(k\tau_0)$ is the $l$th spherical Bessel function. When $k\tau_0\ll1$, the Bessel function reads $j_1(k\tau_0)\sim1/(\sqrt{2}k\tau_0)$. On the other hand, modes that re-entered the horizon near the matter-radiation equality and sometime around the completion of reheating contribute to the overall transfer function through the factors (which can be regarded as transfer functions themselves)
\begin{equation}
    \mathcal{T}_1^2\left(\frac{k}{k_\text{eq}}\right)= 1+0.57\left(\frac{k}{k_\text{eq}}\right)+3.42\left(\frac{k}{k_\text{eq}}\right)^2\,,\label{eq: transfunc1GR}
\end{equation}
and
\begin{equation}
    \mathcal{T}_2^2\left(\frac{k}{k_\text{rh}}\right)=\left[ 1-0.22\left(\frac{k}{k_\text{rh}}\right)^{3/2}+0.65 \left(\frac{k}{k_\text{rh}}\right)^2\right]^{-1} \,.
\end{equation}
For simplicity, the late-time entropy production is not considered. It is worth noting that the coefficients of the transfer function for modes that entered the horizon during the early MD-like phase could vary depending on the inflationary model under consideration (the bulk of this work will play in this wiggle room).

Substituting \eqref{eq: infPt} and \eqref{eq: transF} into \eqref{eq: Pt}, then in \eqref{eq: OmGW}, we obtain
\begin{align}\label{eq: GWspectrum}
    \Omega_{\rm GW}h_0^2 = \frac{\Omega_m^2h_0^2}{96\pi^2H_0^2f^2\tau_0^4}\mathcal{T}_1^2\left(\frac{f}{f_\text{eq}}\right)\mathcal{T}_2^2\left( \frac{f}{f_\text{rh}}\right)r \mathcal{P}_S\left(\frac{f}{f_\ast}\right)^{n_T}\,,
\end{align}
where $\tau_0\sim2 H_0^{-1}$ and $k =2\pi f$. This is the form of the spectrum that we shall mainly consider in this work; $n_T$, $r$, and $T_\text{rh}$ are free parameters to be constrained by the experiment.

\subsection{PTA constraints on standard inflation}
\label{subsec:pta_cosntraints_standard_inflation}

We emphasize that the results of this section are not new and have been thoroughly discussed in, e.g., \cite{NANOGrav:2023hvm}, but we include it for completeness and in preparation for the main part of the work that is to follow. We hope that our independent perspective also adds some value.

Before we present the PTA constraints on inflationary GWs, we keep in mind important observational constraints with other cosmological probes. First of all, the reheating temperature, $T_\text{rh}\gtrsim 10\,\text{MeV}$, must not spoil BBN \cite{Allen:1996vm, Maggiore:1999vm, Cooke:2013cba}.
To satisfy the CMB constraints \cite{Planck:2018jri, Planck:2018vyg}, we consider an upper bound to the tensor-to-scalar ratio $r\leq 0.036$. Keeping this information in mind, we now show the constraints obtained by fitting \eqref{eq: GWspectrum} to the nanohertz GW spectrum observed by PTAs, naturally supporting a blue-tilted tensor power spectrum, or $n_T>0$.  

\begin{figure}[!ht]
    \centering
    \includegraphics[width=0.7\linewidth]{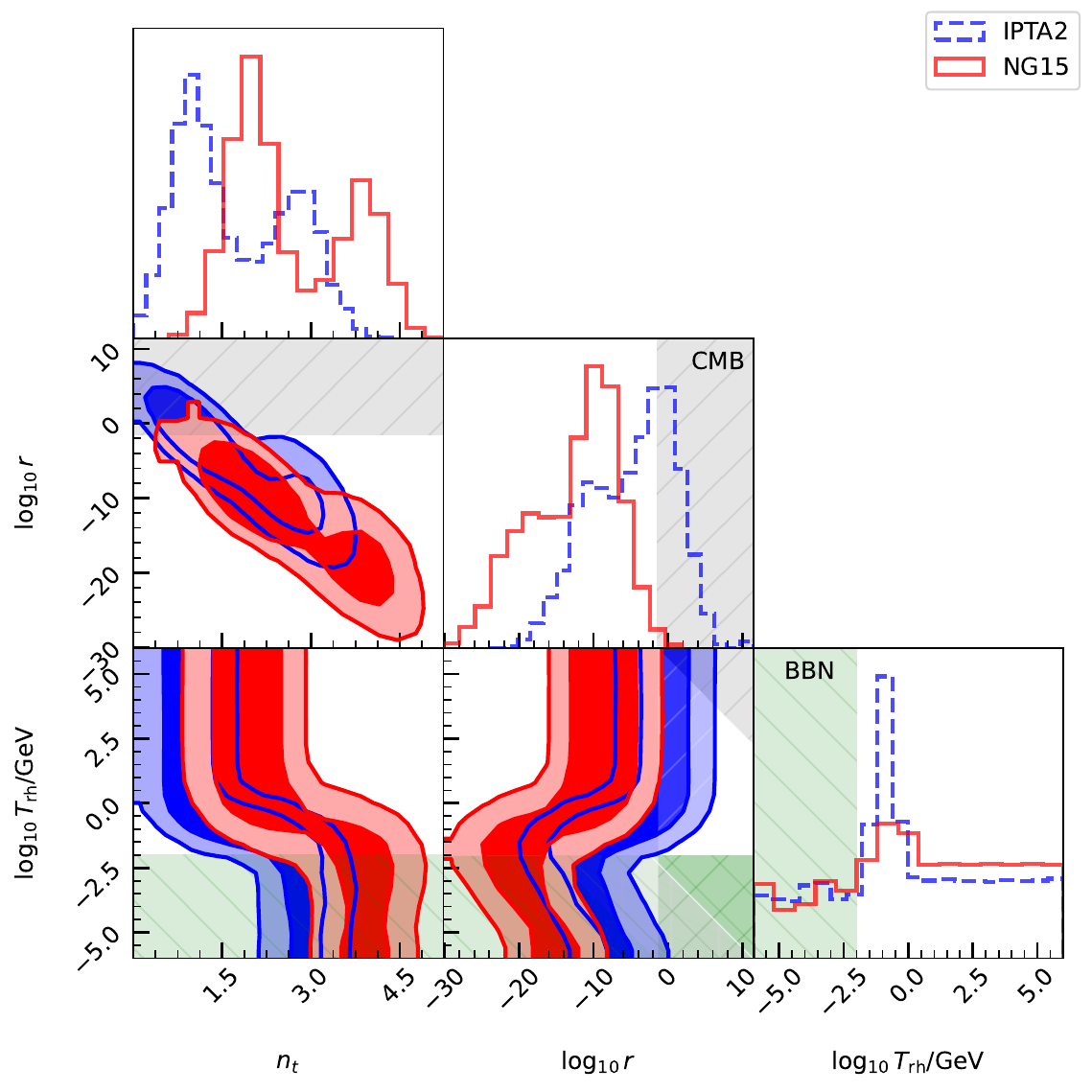}
    \caption{
    Standard (GR) inflation---Constraints (posteriors) of inflationary parameters $n_T$, $r$, and reheating temperature $T_{\rm rh}$ obtained with the NANOGrav 15 years data (red contours) \cite{NANOGrav:2023hvm} and the IPTA-DR2 data set (blue contours) \cite{Antoniadis:2022pcn}. The contours show $68\%$ and $95\%$ confidence levels. The gray and green regions are excluded by CMB and BBN observations, respectively.
    }
    \label{fig:corner_gr_inflation}
\end{figure}

\begin{figure}[!ht]
    \centering
    \includegraphics[width=0.495\linewidth]{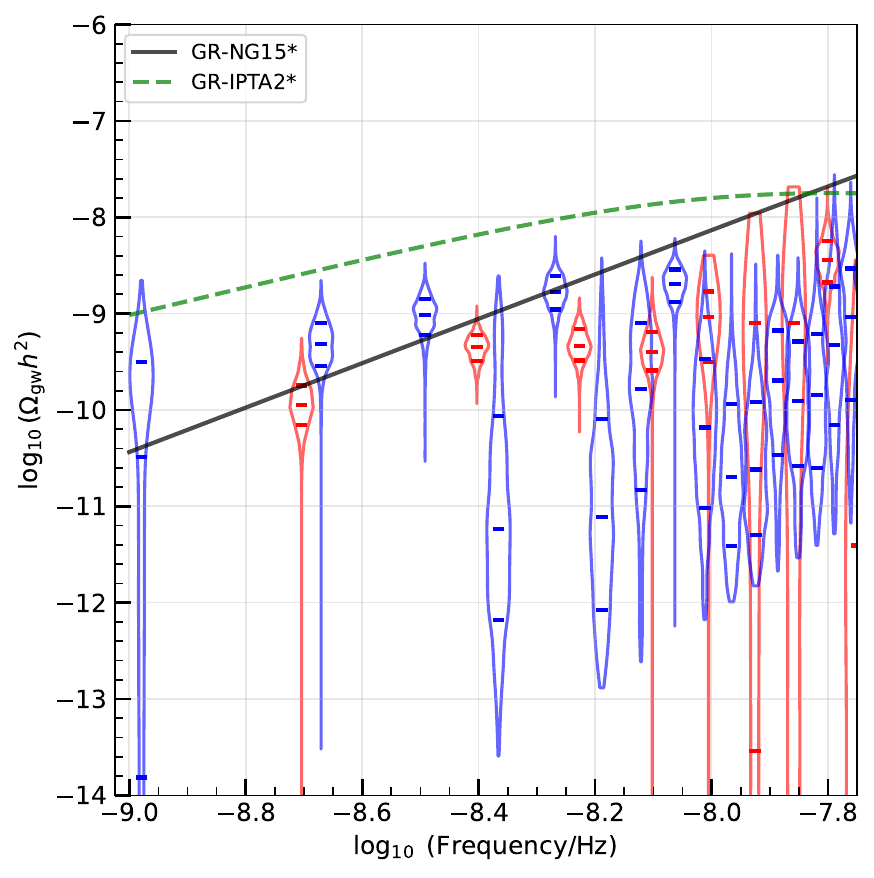}
    \includegraphics[width=0.495\linewidth]{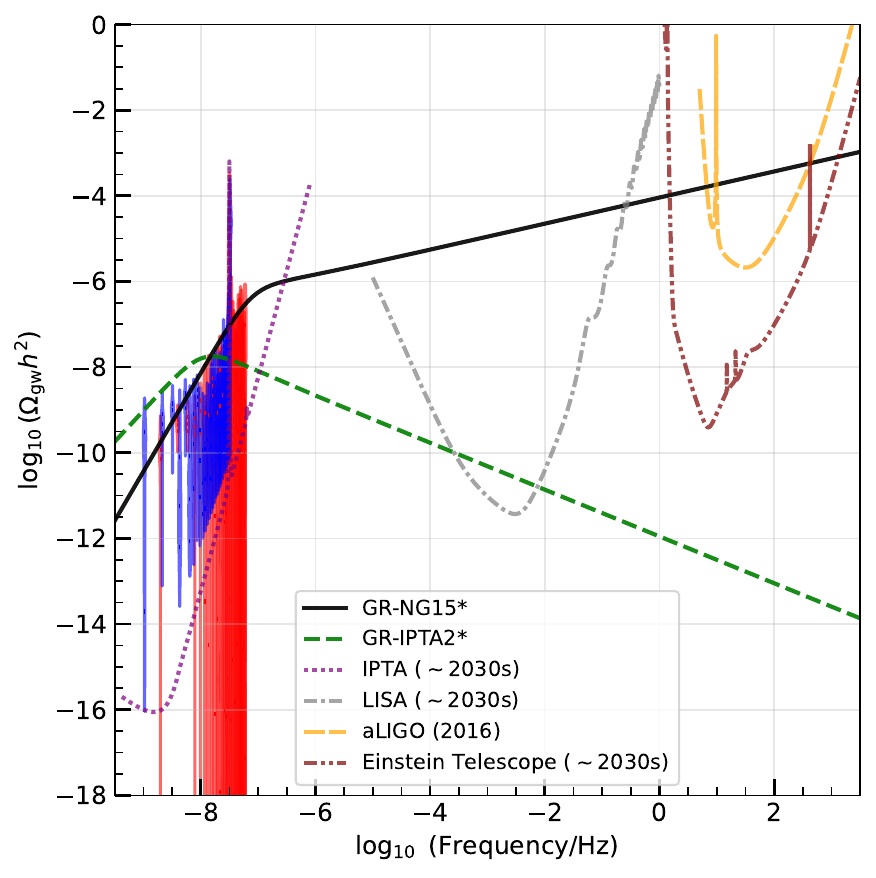}
    \caption{Standard (GR) inflation---[left panel] Best fit curves (bands) of the GW spectrum obtained with the NANOGrav 15 years data (NG15) \cite{NANOGrav:2023hde} and the IPTA-DR2 data set (IPTA2) \cite{Antoniadis:2022pcn}; [right panel] best-fit curves to nanohertz data and GW instrument sensitivity curves (aLIGO, ET, LISA, IPTA/SKA) in other bands, created using \texttt{gwent} \cite{Kaiser:2020tlg}. Red violins are free spectra fits by NANOGrav, blue violins by IPTA-DR2.}
    \label{fig:violins_gr_inflation}
\end{figure}

We investigated the joint parameter space $(r, n_T, T_{\rm rh})$ of the tensor-to-scalar ratio, the tensor spectral tilt, and the reheating temperature. In other words, we keep the three as free parameters and let the observational data determine their acceptable values. For this purpose, we consider the NANOGrav 15-year data set (NG15) \cite{NANOGrav:2023hde} and the IPTA Data Release 2 (IPTA2) \cite{Antoniadis:2022pcn}, utilizing the public code \texttt{PTArcade} \cite{Mitridate:2023oar} to perform the Bayesian spectral fit to PTA data \cite{Lamb:2023jls}. We consider flat priors for each of the parameters: $n_T\in [0,6]$,  $\log_{10}r\in [-40, 20]$, and $\log_{10}T_\text{rh}\in [-6, 6]$. Part of the prior space goes outside permitted CMB and BBN bounds \cite{Cooke:2013cba, Planck:2018vyg}, which we opt to consider post-PTA analysis. It is worth noting that other PTA data sets are consistent with one another, and can also be considered. We use a default \texttt{PTArcade} configuration with the Hellings and Downs correlation \cite{Hellings:1983fr, Domenech:2024pow} considered, 14 SGWB components, 30 red noise components, and $n_{\rm steps}=3\times10^6$ samples. Our results are robust to changes to this setup.

Figures ~\ref{fig:corner_gr_inflation} and \ref{fig:violins_gr_inflation} show the results of our analysis, with constraints $\log_{10} r = -11.4^{+4.4}_{-8.3}$, $n_T=2.3^{+1.6}_{-0.6}$, and $\log_{10} T_{\rm rh}/{\rm GeV} = 0.3^{+3.9}_{-3.5}$ from the NANOGrav data set \cite{NANOGrav:2023hvm} and $\log_{10} r = -4.0^{+4.8}_{-7.3}$, $n_T=1.5^{+1.4}_{-0.7}$, and $\log_{10} T_{\rm rh}/{\rm GeV} = -0.5^{+4.3}_{-2.8}$ from the IPTA data. This shows that PTA data favor inflationary models with a blue-tilted spectrum for the tensor perturbations, contradictory to standard single-field inflation models in GR that produce a red-tilted spectrum ($n_T < 0$). Moreover, the right panel of Figure~\ref{fig:violins_gr_inflation} shows that the blue-tilted spectrum of GWs produced from inflationary models could, in principle, be constrained by current and future space- and ground-based interferometric experiments when $n_T\simeq 2$. 

{
It is worth highlighting a mild tension between the PTA data sets, in this case between the older IPTA Data Release 2 and the more recent NANOGrav 15 year data set. The spectrum plots (figures 2 and 4) show this tension; the IPTA DR2 data has a systematically higher spectra compared to the NANOGrav data, although mildly since the error bars are huge. Nonetheless, we can expect that the upcoming IPTA Data Release 3 \cite{InternationalPulsarTimingArray:2023mzf} will be free from this tension, and remain consistent with CMB bounds.
}

In standard potential-driven single-field inflation, the tensor spectral index is obtained as $n_T=-2\epsilon$, where $\epsilon \equiv M_{\rm pl}^2 V_{,\phi}^2/(2V^2) \ll 1$ is the so-called slow-roll parameter and $V, V_{, \phi}$ are the inflaton potential and its derivative. Since $\epsilon$ is always positive for canonical models, $n_T$ always turns out to be negative. Similarly, red-tilted spectra are also produced in potential-driven G-inflation models \cite{Kamada:2012se}. 
This suggests that a red-tilted spectrum ($n_T < 0$) in GR and G-inflation is a likely scenario. In contrast, models with a GB coupling are well established to produce a blue-tilted GW spectrum \cite{Koh:2016abf, Koh:2018qcy, Mishima:2019vlh}, aligned with PTA data. This brings us to the rest of this work, which is studying the implications of PTA data on GB inflation.

\section{Gravitational Waves from Gauss-Bonnet Inflation} \label{sec: 3}

In this section, we discuss in detail the mechanism for the production of GWs in GB inflation.

\subsection{Background dynamics}
\label{subsec:background_dynamics}
The action for GB inflation is given by~\cite{Satoh:2008ck, Satoh:2010ep, Guo:2010jr, Jiang:2013gza, Koh:2014bka, vandeBruck:2016xvt, Bhattacharjee:2016ohe, Koh:2016abf, Nozari:2017rta, Wu:2017joj, Nojiri:2017ncd, Koh:2018qcy, Chakraborty:2018scm, Tumurtushaa:2018lnv, Mishima:2019vlh, Koh:2023zgn, Odintsov:2023weg}
\begin{align}\label{eq: fullaction}
    S&=\int d^4 x \sqrt{-g}\left[\frac{M_{\rm pl}^2}{2}R -\frac12 g_{ab}\nabla^a \phi\nabla^b \phi- V(\phi) - \frac{1}{2}\xi(\phi) R_{\rm GB}^2 \right]+S_{m,r}\,,
\end{align}
where $R_{\rm GB}^2=R_{\mu\nu\rho\sigma}R^{\mu\nu\rho\sigma}-4R_{\mu\nu}R^{\mu\nu}+R^2$ is the GB term. The action involves the Einstein-Hilbert term and a canonical scalar field with the potential $V(\phi)$, which couples to the GB term through the coupling function $\xi(\phi)$. The $S_{m,r}$ denotes the action for matter and radiation. 
In a spatially flat FLRW metric, 
\begin{align}\label{eq:flatmetric}
    ds^2=g_{\mu\nu}dx^\mu dx^\nu=-dt^2 +a(t)^2\delta_{ij}dx^idx^j\,,
\end{align}
the background equations of motion (obtained by varying  \eqref{eq: fullaction} with respect to $g_{\mu\nu}$ and $\phi$) can be shown to be 
\begin{subequations}\label{eq: bEOMs}
\begin{align}\label{eq: bEOM1}
    3M_{\rm pl}^2 H^2 &= \frac{1}{2}\dot{\phi}^2 + V +12\dot{\xi}H^3\,,\\
    M_{\rm pl}^2(2\dot{H}+3H^2) &= -\frac{1}{2}\dot{\phi}^2 +V +4\ddot{\xi}H^2+8\dot{\xi}H\left(\dot{H}+H^2\right)\,,\label{eq: bEOM2}\\
    \ddot{\phi} +3H\dot{\phi} +V_{,\phi} &=-12\xi_{,\phi}H^2\left(\dot{H}+H^2\right)\,,\label{eq: bEOM3}
\end{align}
\end{subequations}
where $H\equiv \dot{a}/a$ is the Hubble function. The overdot denotes a time derivative and $\xi_{, \phi}=d\xi/d\phi$. For constant $\xi(\phi)$, the equations of motion reduce to the standard case (GR with canonical scalar field), owing to the GB term becoming a mere boundary term. Nontrivial imprints of the GB term are tied to a varying GB coupling function $\xi(\phi) \neq\text{constant}$. This is the focus of this work.

In potential-driven inflation, the scalar field is considered to slowly roll along a nearly flat potential and to be headed towards a minimum. With a GB coupling, the potential and friction terms in \eqref{eq: bEOMs} dominate, as reflected in the following slow-roll approximations: 
\begin{align}\label{eq: slconGR}
    \dot{\phi}^2\ll V \,, \quad 
    \ddot{\phi} \ll H \dot{\phi}\,, \quad  4\dot{\xi}H\ll M_{\rm pl}^2\,.
\end{align}
The last one is due to the GB term, which is considered a small correction to GR. Consequently, (\ref{eq: bEOM1}) and (\ref{eq: bEOM3}) read
\begin{subequations}\label{eq: sreqs}
   \begin{align}
    &3M_{\rm pl}^2H^2\simeq V\,,\\
    &3H\dot{\phi}+V_{,\phi} \simeq -\frac{4}{3M_{\rm pl}^4}V^2 \xi_{,\phi}\,,
    \end{align} 
\end{subequations}
where we assumed $\xi_{,\phi}V_{,\phi}\ll M_{\rm pl}^2$.
From \eqref{eq: sreqs}, one obtains 
\begin{align}\label{eq: sqrtbeta}
    \frac{\dot{\phi}}{M_{\rm pl} H}\simeq -\frac{M_{\rm pl}}{V}\left( V_{,\phi}+\frac{4}{3 M_{\rm pl}^4}\xi_{,\phi}V^2\right)\,.
\end{align}
To yield the conditions in \eqref{eq: slconGR}, the following inequalities must hold during slow-roll inflation~\cite{Satoh:2010ep}
\begin{align}
    \frac{M_{\rm pl}^2}{V^2}Q^2\ll1\,, \quad \frac{M_{\rm pl}^2 V_{,\phi}}{2V^2}Q-\frac{M_{\rm pl}^2}{V}Q_{,\phi}\ll 1\,,\quad \frac{M_{\rm pl}^2 V_{,\phi}}{V^2}Q - \frac{M_{\rm pl}^2}{V^2}Q^2\ll1\,,
\end{align}
where 
\begin{align}\label{eq: forQ}
    Q\equiv V_{,\phi}+\frac{4}{3M_{\rm pl}^4}\xi_{,\phi}V^2\,.
\end{align}
In the trivial limit, $\xi(\phi) \sim \text{constant}$, the above equations reduce to standard slow-roll inflation. The stability of de Sitter solutions in \eqref{eq: fullaction} can be assessed by referring to an effective potential \cite{Skugoreva:2014gka, Pozdeeva:2019agu, Pozdeeva:2020apf, Vernov:2021hxo, Pozdeeva:2024ihc, Khan:2022odn, Gangopadhyay:2022vgh, Yogesh:2024mpa, Yogesh:2025wak}
\begin{align}\label{eq: effpot}
    \frac{V_{\rm eff}}{M_{\rm pl}^4}=-\frac{1}{4}\left(\frac{M_{\rm pl}^4}{V} - \frac{4}{3}\xi \right)\,,
\end{align}
and its derivative
\begin{align}\label{eq: forVeff}
    \frac{V_{{\rm eff},\phi}}{M_{\rm pl}^3}=\frac{1}{4}\left(\frac{M_{\rm pl}^4}{V}\right)^2\left[\frac{V_{,\phi}}{M_{\rm pl}^3}+\frac{4}{3}(M_{\rm pl}\xi_{,\phi})\left(\frac{V}{M_{\rm pl}^4}\right)^2\right]=\frac{1}{4}\frac{Q/M_{\rm pl}^3}{\left(V/M_\text{pl}^4\right)^2}
    \,.
\end{align}
However, this is not well defined for $V(\phi)=0$ for which inflationary scenarios have been identified as unstable~\cite{Hikmawan:2015rze}. Also, we assume $V(\phi)>0$ during inflation. Following \cite{Satoh:2010ep}, it is useful to define the following parameters~\footnote{Our definitions of the slow-roll parameters are consistent with those in \cite{Koh:2014bka} up to $\alpha=\epsilon$, $\sigma=\delta_1/4$, $2\beta= 2\epsilon-\delta_1$.}
\begin{align}\label{eq: srparams}
    \alpha\equiv \frac{M_{\rm pl}^2}{2}\frac{V_{,\phi}}{V^2}Q =-\frac{\dot{H}}{H^2}
    \,,\quad 
    \beta \equiv  \frac{M_{\rm pl}^2}{2}\frac{Q^2}{V^2}=\frac{\dot{\phi}^2}{2M_{\rm pl}^2 H^2}
    \,,\quad \gamma \equiv M_{\rm pl}^2 \frac{Q_{,\phi}}{V}\,,\quad \sigma \equiv\frac{H\dot{\xi}}{M_{\rm pl}^2}
    \,.
\end{align}
In the case of slow-roll inflation, these parameters are assumed to be small during inflation. 
The amount of inflation is quantified in the number $N$ of $e-$folds, which is obtained in our model as
\begin{align}\label{eq: efold}
    N=\int H dt = \int^{\phi_\ast}_\phi d\phi \frac{H}{\dot{\phi}}= \frac{1}{M_{\rm pl}^2} \int_{\phi_\ast}^\phi d\phi\frac{V}{Q}
    \,,
\end{align}
where $\phi_{\ast}$ denotes the scalar field value at the horizon-crossing time of a mode with $k=k_\ast$. Having reviewed the background field equations and the slow-roll approximation in GB inflation, we are now prepared to derive equations of motion for scalar and tensor perturbations and obtain observable quantities. No vector modes play an important role since there is no source for vectors at the linear perturbation level. In computing the observable quantities, we regard the parameters in (\ref{eq: srparams}) as constants at the time of horizon crossing but not necessarily small. 

\subsection{Scalar perturbations}
In the comoving gauge, in which $\delta\phi=0$, the scalar perturbation of the linearized metric takes the following form
\begin{align}
    ds^2 = a(\tau)^2\left[ -d\tau^2 +\left( 1-2\mathcal{R}\right)\delta_{ij}dx^idx^j\right]\,,
\end{align}
where $\mathcal{R}$ represents the curvature perturbation on the uniform field hypersurfaces. The conformal time $\tau$ is related to the physical time $t$ through $d\tau \equiv dt/a(t)$. Expanding $\mathcal{R}$ in Fourier modes,
\begin{align}
    \mathcal{R}(\tau, \textbf{x}) = \int \frac{dk^3}{(2\pi)^{3/2}}\mathcal{R}_\textbf{k}(\tau) e^{i\textbf{k}\cdot\textbf{x}}\,,
\end{align}
the action for $\mathcal{R}$ becomes~\cite{Satoh:2010ep}
\begin{align}\label{eq: pertS}
    S=\frac{1}{2}\int d\tau\int\frac{dk^3}{(2\pi)^3}z_S^2\left(|\mathcal{R}'_\textbf{k}|^2-c_S^2k^2|\mathcal{R}_\textbf{k}|^2 \right)\,,
\end{align}
where 
\begin{align}\label{eq: zs}
    &z_S^2\equiv a^2 \left(\frac{1-4\sigma}{1-6\sigma}\right)^2\left( \frac{\phi'^2}{M_{\rm pl}^2\mathcal{H}^2}+\frac{24\sigma^2}{1-4\sigma}\right)\simeq 2a^2\beta\,,\\
    &c_S^2\equiv 1+\frac{2a^2}{z_S^2}\left(\frac{4\sigma}{1-6\sigma}\right)^2\left[\left(\frac{\mathcal{H}'}{\mathcal{H}^2}-1\right)\left( 1-4\sigma\right)-\sigma+\frac{\xi''-\mathcal{H}\xi'}{M_{\rm pl}^2a^2}\right]\simeq 1\,.\label{eq: cs}
\end{align}
The approximate equalities in \eqref{eq: zs} and \eqref{eq: cs} are derived up to the first order in the slow-roll parameter $\sigma$ in \eqref{eq: srparams}. By defining a new variable, $v_S=z_S \mathcal{R}_\textbf{k}$, one can rewrite the action in terms of this newly defined variable as  
\begin{align}\label{eq: perteqS}
    S = \frac{1}{2}\int d\tau \int \frac{dk^3}{(2\pi)^3}\left(|v_S'|^2- c_S^2k^2|v_S|^2+\frac{z_S''}{z_S}|v_S|^2 \right)\,.
\end{align}
Consequently, the perturbed equations of motion for $v_S$ can be obtained as
\begin{align}
    v_S'' + \left(c_S^2k^2-\frac{z_S''}{z_S} \right)v_S=0\,.
\end{align}
From (\ref{eq: zs}) with $\sigma\ll1$, the effective potential can be obtained as
\begin{align}
    \frac{z_S''}{z_S}=\frac{\nu_S^2-1/4}{\tau^2}\,,
\end{align}
where
\begin{align}
    \nu_S = \frac{3}{2}+\frac{3\alpha-\gamma}{1-\alpha}\,.
\end{align}
Imposing the Bunch-Davies vacuum for the initial condition at $\tau\rightarrow-\infty$, we can solve \eqref{eq: perteqS} to obtain
\begin{align}\label{eq: solS}
    v_S = \frac{\sqrt{-\pi \tau}}{2}e^{i\frac{\pi}{2}\left(\frac{1}{2}+\nu_S\right)}H_{\nu_S}^{(1)}(-k\tau)\,,
\end{align}
where $H_{\nu_S}^{(1)}(-k\tau)$ is the Hankel function of the first kind, in the form
\begin{align}\label{eq: Hankel}
    H^{(1)}_{\nu_S}(x)\simeq \frac{2}{1-e^{2i\pi\nu_S}}\left[\frac{1}{\Gamma(1+\nu_S)}\left(\frac{x}{2}\right)^{\nu_S}-\frac{e^{i\pi\nu_S}}{\Gamma(1-\nu_S)}\left(\frac{x}{2}\right)^{-\nu_S}\right]\,.
\end{align}
Consequently, the power spectrum of the scalar perturbation for super-horizon modes is calculated with \eqref{eq: solS} on the large scales, leading to
\begin{align}\label{eq: PowerS}
    \mathcal{P}_S = \frac{k^3}{2\pi^2M_{\rm pl}^2}\left|\frac{v_S}{z_S}\right|^2\simeq \frac{\left(1-\alpha\right)^2}{8\pi^2\beta}\left(\frac{\mathcal{H}}{M_{\rm pl} a}\right)^2\left( \frac{-k \tau}{2}\right)^{3-2\nu_S}\,,
\end{align}
where we assumed $\sigma\ll1$.
The scalar spectral index $n_S$ becomes
\begin{align}\label{eq: nSm1}
    n_S-1 = 3-2\nu_S=-\frac{6\alpha-2\gamma}{1-\alpha}\,.
\end{align}
In the limit of a vanishing GB coupling, that is, $\xi \rightarrow 0$, the slow-roll parameters $\alpha$ and $\gamma$ coincide with the standard scenario. 

\subsection{Tensor perturbations}
The tensor perturbation of the spacetime metric is given by \eqref{eq: metricT}; and in Fourier space, $h_{ij}(\tau, \textbf{x})$, by \eqref{eq: fourierTrans}. By substituting the metric into \eqref{eq: fullaction}, we obtain the quadratic action for $h_\mathbf{k}^A(\tau)$ as follows
\begin{align}\label{eq:actionforhk}
S=\frac{1}{2}\int d\tau\int\frac{d^3k}{(2\pi)^3}a^2\left[\left(1-\frac{4\mathcal{H}\xi'}{M_{\rm pl}^2a^2}\right)|h_\mathbf{k}' |^2 -k^2\left(1+\frac{4\mathcal{H}\xi'}{M_{\rm pl}^2a^2}-\frac{4\xi''}{M_{\rm pl}^2a^2}\right)|h_\mathbf{k}|^2\right]\,.
\end{align}
Consequently, the perturbed equations of motion can be obtained as
\begin{align}\label{eq: waveeq}
    h_{\mathbf{k}}'' + \frac{z_T'}{z_T} {h_{\mathbf{k}}}' + k^2 c_T^2 h_{\mathbf{k}}=0\,,
\end{align}
where
\begin{align}\label{eq:zandct}
    z_T^2\equiv a^2\left(1-\frac{4\mathcal{H}\xi'}{M_{\rm pl}^2a^2}\right)=a^2(1-4\sigma)\,, \qquad c_T^2\equiv 1+\frac{8\mathcal{H}\xi'}{M_{\rm pl}^2 z_T^2}-\frac{4\xi''}{M_{\rm pl}^2 z_T^2}\simeq 1+4\sigma \,.
\end{align}
This equation is the GB extended version of \eqref{eq: hGR} and can also be written as 
\begin{align}
    v_T'' + \left(c_T^2k^2-\frac{z_T''}{z_T} \right)v_T=0\,.
\end{align}
From (\ref{eq:zandct}) with $\sigma\ll1$, the effective potential can be written as
\begin{align}\label{eq: perteqT}
    \frac{z_T''}{z_T}=\frac{\nu_T^2-1/4}{\tau^2}\,,
\end{align}
where
\begin{align}
    \nu_T\equiv \frac{3}{2}+\alpha\,.
\end{align}
Imposing the Bunch-Davies vacuum for the initial condition at $\tau\rightarrow-\infty$, we can solve \eqref{eq: perteqT} and obtain the solution
\begin{align}\label{eq: solT}
    v_T = \frac{\sqrt{-\pi \tau}}{2}e^{i\frac{\pi}{2}\left(\frac{1}{2}+\nu_T\right)}H_{\nu_T}^{(1)}(-k\tau)\,.
\end{align}
The form of $H_{\nu_T}^{(1)}(-k\tau)$ is given in \eqref{eq: Hankel}. The power spectrum of the tensor mode is calculated with \eqref{eq: solT} on the large scales, leading to the expression~\cite{Satoh:2010ep}
\begin{align}\label{eq: Ppol}
    \mathcal{P}_T = \frac{2k^3}{\pi^2 M_{\rm pl}^2}\left|\frac{v_T}{z_T}\right|^2\simeq \frac{2\left(1-\alpha\right)^2}{\pi^2}\left(\frac{\mathcal{H}}{M_{\rm pl} a}\right)^2\left( \frac{-k \tau}{2}\right)^{3-2\nu_T}\,,
\end{align}
where we only assumed $\sigma\ll1$.
The tensor spectral index $n_T$ becomes
\begin{align}\label{eq: nT}
    n_T = 3-2\nu_T=-2\alpha\,.
\end{align}
Once again, in the limit $\xi\rightarrow0$, $n_T\sim-2\epsilon$ is realized, where $\epsilon= M_{\rm pl}^2V_{,\phi}^2/(2V^2)$ is the slow-roll parameter of the GR limit. The tensor power spectrum is blue-tilted if $n_T>0$ and red-tilted if $n_T<0$. The observation of a blue-tilted signal can be considered as evidence in favor of our model. The tensor-to-scalar ratio is given by
\begin{align}\label{eq: r}
    r\equiv \frac{\mathcal{P}_T}{\mathcal{P}_S}\simeq 16 \beta\,.
\end{align}
We will use this, along with other observable quantities such as $n_T$ and $n_S$, in the following sections to constrain and study GB inflation with PTA data.

{
It is worth clarifying that the tensor-to-scalar ratio, and the tensor and scalar spectral indices (Eqs.~(\ref{eq: nSm1}), (\ref{eq: nT}), and (\ref{eq: r})) include the GB corrections. This can be realized in the limit of vanishing coupling ($\xi \rightarrow 0$); in this case, the $\xi$-dependent term in Eq.~(\ref{eq: forQ}) vanishes, leaving us with $Q=V_{,\phi}$. Then, the slow-roll parameters in Eqs.~(\ref{eq: srparams}) reduce to the standard GR scenario.
}

\subsection{Nanohertz GWs in GB Inflation: PTA Constraints}
The inflationary GW spectrum for our model remains to be described by \eqref{eq: GWspectrum}. However, because of the GB coupling, GW modes evolve differently inside the horizon compared to GR, changing the transfer functions. The relevant transfer functions for our model are calculated by numerically integrating \eqref{eq: waveeq}. In integrating the equation for our model, we may assume that the effects of the GB term are negligible by the time of matter-radiation equality~\cite{Koh:2023zgn}. Thus, the transfer function for modes that entered the horizon sometime before and after the matter-radiation equality is the same as (\ref{eq: transfunc1GR}).
On the other hand, for the modes that re-entered the horizon after the end of inflation but before the completion of reheating, the transfer function should reflect the GB effects. In this case, the transfer function can be approximated with the following form~\cite{Koh:2018qcy}
\begin{align}\label{eq:transF2GB}
\mathcal{T}_2^2\left(\frac{k}{k_\text{rh}}\right)=\left[1+c_1 \left(\frac{k}{k_\text{rh}}\right)^{\frac{3}{2}}+c_2\left(\frac{k}{k_\text{rh}}\right)^2\right]^{-1}\,,
\end{align}
where the $c_1$ and $c_2$ depend on the form GB coupling function $\xi(\phi)$; in particular, one needs to solve the background equations of motion in (\ref{eq: bEOMs}) for given potential and GB coupling functions and use the result to integrate \eqref{eq: waveeq}, see Ref.~\cite{Koh:2018qcy} for an example. Since the form of GB coupling is not fixed in the present work, we regard the $c_1$ and $c_2$ coefficients as free parameters of the model and let the observational data constrain their values, by fitting the GW spectrum of our model in \eqref{eq: fullaction} to PTA data using \eqref{eq: OmGW} and (\ref{eq: nT}--\ref{eq:transF2GB}). 
\begin{table}[!ht]
    \centering
        \caption{Constraints on the model parameters in light of the NANOGrav 15 years data (NG15) \cite{NANOGrav:2023hde} and the IPTA Data Release 2 (IPTA2) \cite{Antoniadis:2022pcn}.}
    \begin{tabular}{c|c|c}
        \hline \hline
        Parameters & NG15 & IPTA-DR2 \\
        \hline 
        \rule{0pt}{3ex} $n_T$ & $2.08^{+ 1.11}_{ - 0.41}$   & $1.14^{+ 1.33}_{ - 0.5}$ \\
        \rule{0pt}{3ex} $\log_{10}r$   & $-9.54^{+ 3.21}_{- 5.37}$ & $-1.69^{+ 3.8}_{ - 7.1}$ \\
        \rule{0pt}{3ex} $\log_{10} T_{\rm rh}/{\rm GeV}$ & $2.34^{+ 2.48}_{ - 3.02}$  & $1.7^{+2.93}_{ - 2.69}$ \\
        \rule{0pt}{3ex} $c_1$ & $0.78^{+ 6.32}_{ - 6.94}$  & $0.88^{+ 6.28}_{ - 6.86}$\\
        \rule{0pt}{3ex} $c_2$ & $1.94^{ + 5.54}_{ - 7.65}$  & $2.58^{+ 5.19}_{ - 7.77}$ \\
        \hline \hline
    \end{tabular}
    \label{tab:1}
\end{table}

\begin{figure}[!ht]
    \centering
    \includegraphics[width=0.9\linewidth]{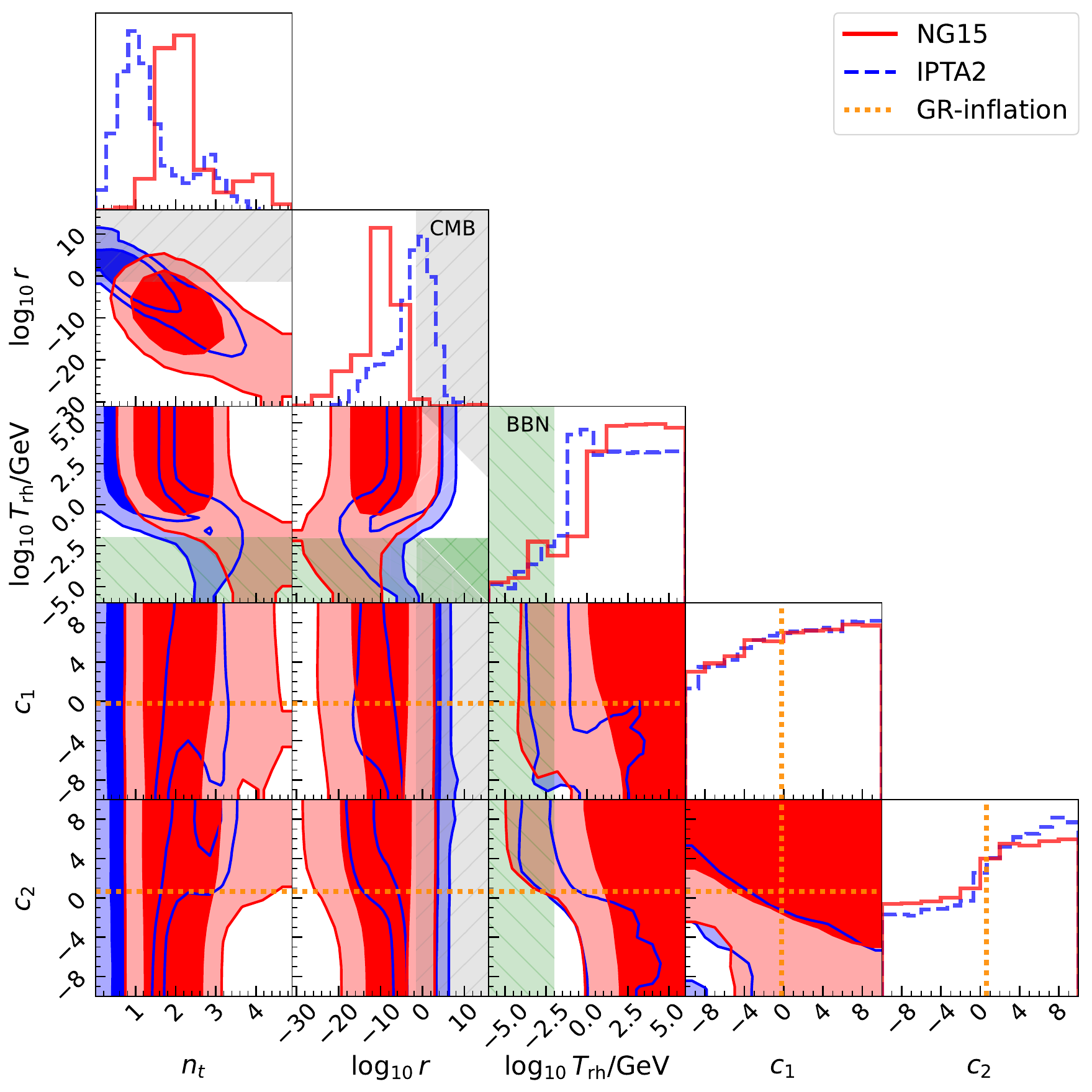}
    \caption{Gauss-Bonnet inflation---Constraints (posteriors) of inflationary parameters $n_T$, $r$, reheating temperature $T_{\rm rh}$, and Gauss-Bonnet couplings ($c_1$ and $c_2$) obtained with the NANOGrav 15 years data (NG15, in red) \cite{NANOGrav:2023hde}  and the IPTA-DR2 data set (IPTA2, in blue) \cite{Antoniadis:2022pcn}. The contours show $68\%$ and $95\%$ confidence intervals. The corresponding marginalized constraints are provided in Table~\ref{tab:1}. The gray and green regions are excluded by CMB and BBN observations, respectively. {The orange dotted line is shown as a reference to standard inflation when the GB terms vanish.}}
    \label{fig:corner_gb_inflation}
\end{figure}

\begin{figure}[!ht]
    \centering
    \includegraphics[width=0.495\linewidth]{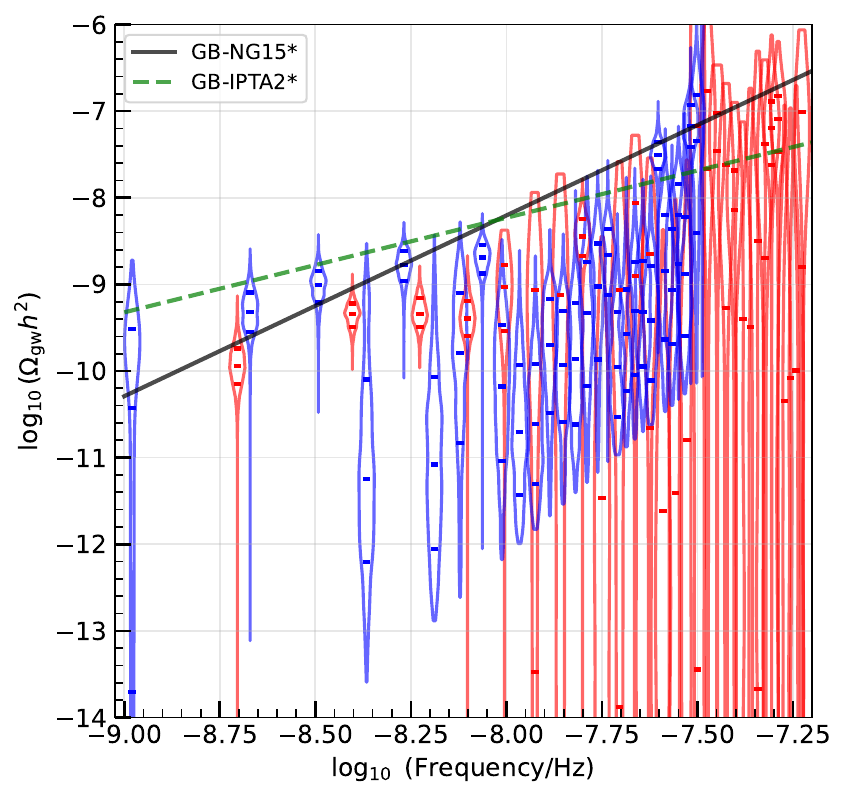}
    \includegraphics[width=0.495\linewidth]{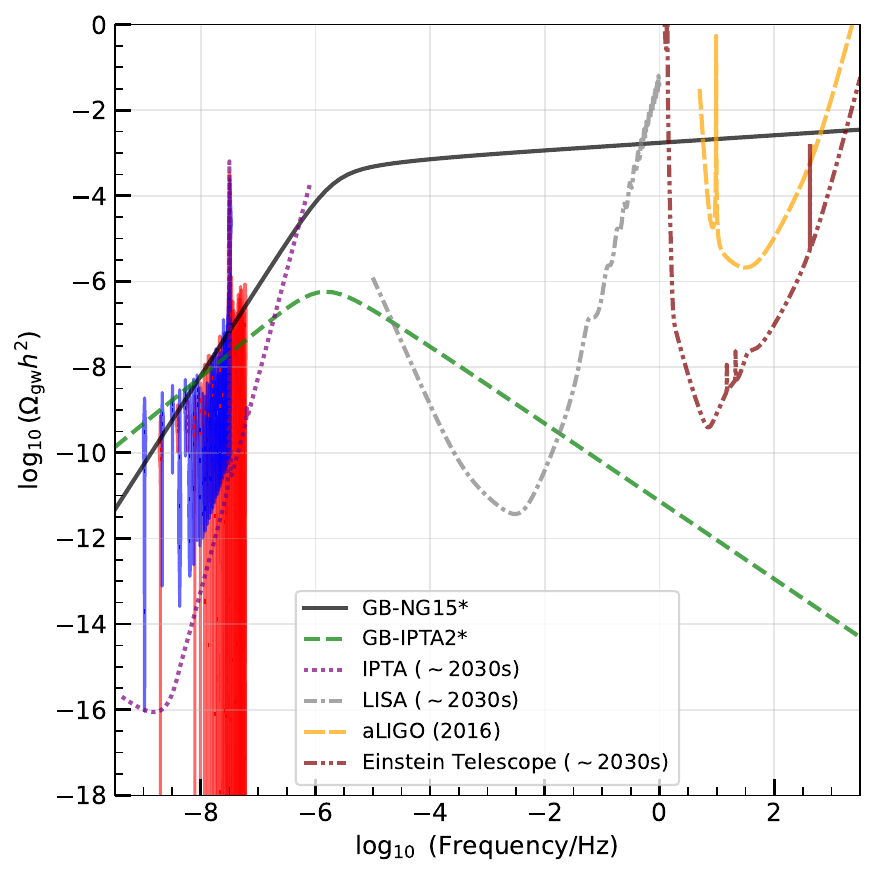}
    \caption{Gauss-Bonnet inflation---[left panel] Best fit curves (bands) of the GW spectrum obtained with the NANOGrav 15 years data \cite{NANOGrav:2023hde} and the IPTA-DR2 data set (IPTA2) \cite{Antoniadis:2022pcn}; [right panel] best-fit curves to nanohertz data and GW instrument sensitivity curves (aLIGO, ET, LISA, IPTA/SKA) in other bands, created using \texttt{gwent} \cite{Kaiser:2020tlg}. Red violins are free spectra fits by NANOGrav, blue violins by IPTA-DR2.}
    \label{fig:violins_gb_inflation}
\end{figure}

To this end, we consider the NANOGrav 15-year data set \cite{NANOGrav:2023hde} and the IPTA Data Release 2 \cite{Antoniadis:2022pcn} free spectra measurements and run the Bayesian fit of our model with the data using \texttt{PTArcade} \cite{Mitridate:2023oar}, as we did in Section \ref{sec: 2} with the same default configuration. We use the same flat priors on $r, n_T$, and $T_{\rm rh}$ as in Section \ref{sec: 2}. In addition, we consider flat priors $c_1, c_2 \in [-10, 10]$ for the GB couplings\footnote{
{
We have also investigated smaller and larger $c_{1,2}$ priors, including $[-5, 5]$ and $[-20, 20]$, but have found no significant addition to the results due to the present PTA sensitivity.
}
}.
Figure~\ref{fig:corner_gb_inflation} and Table \ref{tab:1} (marginalized constraints) present our results on GB inflation. The contours show the confidence intervals $68\%$ and $95\%$ obtained with the NANOGrav \cite{NANOGrav:2023hde} and IPTA \cite{Antoniadis:2022pcn} data. In addition, model-independent BBN ($T_\text{rh}\gtrsim 10\,\text{MeV}$ \cite{Allen:1996vm, Maggiore:1999vm, Cooke:2013cba}) and CMB ($r\leq 0.036$ \cite{Planck:2018jri, Planck:2018vyg}) constraints are shown. Note that there are stronger constraints to the scalar-to-tensor ratio, e.g., $r \leq 0.032$ \cite{Tristram:2021tvh} and $r \leq 0.028$ \cite{Galloni:2024lre}, and our conclusions are insensitive to these choices.

However, while with a slight preference for positive values, we must emphasize that the PTA constraints on the GB coupling through $c_1$ and $c_2$ are, at best, weak. We will return to explain this shortly.

The results imply that the IPTA-DR2 data set prefers a relatively smaller value of the tilt, $n_T$, compared to the NANOGrav 15-year data set, while their preference for the tensor-to-scalar ratio $r$ is the opposite. Similar results were obtained with standard inflation (Section \ref{sec: 2}). Regardless, neither data set shows any strong constraints on the coefficients $c_1$ and $c_2$. As discussed in Section~\ref{sec: 2}, reheating temperature characterizes the peak frequency of suppression in the GW spectrum that occurs due to the reheating phase. This indicates that the peak frequency is much larger than the frequency range to which the PTA experiments are sensitive. To show this, we plot the mean spectral energy density of the GWs of our model in Figure~\ref{fig:violins_gb_inflation} by using \eqref{eq: OmGW} with (\ref{eq: nT}--\ref{eq:transF2GB}). The red and blue violin plots show the NANOGrav and IPTA data, respectively. The non-suppression of the spectrum in the nanohertz band explains the weak sensitivity of the PTA data to $c_1$ and $c_2$. This behavior holds as long as the reheating temperature is higher than the BBN bound, i.e., $T_\text{rh}\gtrsim 10\,\text{MeV}$. Clearly, this is more satisfied in GB inflation compared to the standard case, as shown in Figure \ref{fig:corner_gb_inflation} (supportive of $T_{\rm rh} \gtrsim 10$ MeV), in contrast to Figure \ref{fig:corner_gr_inflation}. 

In Figure \ref{fig:violins_gb_inflation}, the suppression occurs in the higher frequency range, higher than the PTA range, due to the modes that entered the horizon during the phase of reheating, a phase after inflation and before radiation domination. Although the slope of the spectrum is suppressed in the higher frequency range, space- and ground-based GW interferometer experiments such as LISA, aLIGO, and Einstein Telescope may be able to chip in to constrain the GW signal of GB inflation. The right panel of Figure~\ref{fig:violins_gb_inflation} shows that the spectrum is more suppressed for a smaller $n_T$ value, as favored by IPTA-DR2. Therefore, its detectability is subject to future space-based interferometer experiments like LISA. In contrast, the GW spectrum corresponding to the values $n_T$ preferred by the NANOGrav dataset remains within the sensitivity range of ground-based experiments. From a slow-roll approximation standpoint, values of $n_T$ larger than $2$ are not favored, as the $|\alpha|$ parameter in (\ref{eq: nT}) must be smaller than unity during slow-roll inflation.

{
The numerical values of $c_1$ and $c_2$
in Table~1 do not necessarily correspond to a small
GB coupling, and should be treated as entirely phenomenological. When interpreted in GB inflation, the coefficients can be read as functionals of the GB coupling, i.e., $c_1[\xi(\phi)]$ and $c_2[\xi(\phi)]$. Then, we could use the field equations backward to determine observationally-compatible GB couplings. This is the gist of the next section of our paper. We should also note that the sampled tensor-to-scalar ratio and tensor spectral index are covariant with the rest of the parameters such as $c_{1,2}$; in other words, drawing implications based on the marginalized $(n_t, r)$ uses the information in the full parameter space.
}

The following section discusses the implications of the PTA constraints on GB inflation. 

\section{Implications of PTAs for Gauss-Bonnet Inflation} \label{sec: 4}
Results have shown that the PTA data favor inflationary models with a blue-tilted spectrum ($n_{T}>0$) of primordial tensor perturbations (Figure~\ref{fig:corner_gb_inflation} and Table~\ref{tab:1}), a natural consequence of a nontrivial GB coupling.
We discuss the implications of these findings for GB inflation.

\subsection{Implications on Inflaton Potential, GB Coupling, and Effective Potential}
With $n_T$ and $r$ values determined by observational data, it is possible to reconstruct the inflaton potential, its derivative, and the derivative of the GB coupling function for the CMB scale. For instance, using \eqref{eq: PowerS}, the potential at the horizon crossing can be expressed as
\begin{align}\label{eq: potVal}
    \frac{V}{M_{\rm pl}^4} = 24 \pi^2 \frac{\beta}{\left(1-\alpha \right)^2} \mathcal{P}_S = \frac{3\pi^2}{2}r \left(1+\frac{n_T}{2}\right)^{-2} \mathcal{P}_S\,,
\end{align}
and its numerical value gets
\begin{align*}
    \frac{V}{M_{\rm pl}^4}= 3.09\times 10^{-8}r\left( 1+\frac{n_T}{2}\right)^{-2}\left(\frac{\mathcal{P}_S}{2.09\times 10^{-9}}\right)\,,
\end{align*}
where the value of $r$ and $n_T$ can be taken from Table~\ref{tab:1}. Consequently, the derivative of the potential can be obtained from (\ref{eq: srparams}), (\ref{eq: nT}), and (\ref{eq: r}) as
\begin{align}\label{eq: derV}
    \frac{V_{,\phi}}{M_{\rm pl}^3}
    =-3\sqrt{2}\pi^2 n_T \sqrt{r} \left(1+\frac{n_T}{2}\right)^{-2}\mathcal{P}_S\,,
\end{align}
With the use of Table~\ref{tab:1}, (\ref{eq: derV}) can also be written as
\begin{align}
    \frac{V_{,\phi}}{M_{\rm pl}^3}
    =-1.09\times 10^{-9} n_T\sqrt{r}\left(1+\frac{n_T}{2}\right)^{-2}\left(\frac{\mathcal{P}_S}{2.09\times 10^{-9}}\right)\,,
\end{align}
which is negative because both $n_T$ and $r$ are positive. Thus, $V_{,\phi}<0$ is a necessary condition for achieving a blue-tilted tensor spectrum. This implies that the scalar field must experience a climb up the potential during inflation before rolling down.

Since the dynamical equations depend on the derivative of the GB coupling function, implications on the derivative $\xi_{,\phi}$ can be given. From \eqref{eq: srparams}, we obtain 
\begin{align}
    \frac{\beta}{\alpha}
    =1+\frac{2\sqrt{2}}{3M_{\rm pl}^3}\frac{\sqrt{\beta}}{\alpha}V \xi_{,\phi}\,.
\end{align}
Turning this around lets us write down the derivative of the GB coupling as
\begin{align}\label{eq: derXi}
    M_{\rm pl}\xi_{,\phi} = \frac{3M_{\rm pl}^4}{2\sqrt{2}}\frac{1}{\sqrt{\beta} V}\left(\beta-\alpha\right)  
    =  \frac{1}{4\sqrt{2}\pi^2\mathcal{P}_S} \frac{1}{\sqrt{r}}\left(1+\frac{8n_T}{r}\right)\left(1+\frac{n_T}{2}\right)^{2}\,.
\end{align}
Consequently, its value at the horizon crossing is
\begin{align}
    M_{\rm pl}\xi_{,\phi}=8.56\times 10^6\left(\frac{\mathcal{P}_S}{2.09\times 10^{-9}}\right)^{-1} \frac{1}{\sqrt{r}}\left(1+\frac{8n_T}{r}\right)\left(1+\frac{n_T}{2}\right)^{2}\,.
\end{align}
Unlike the slope of the inflaton potential, the slope of the GB coupling function must be positive ($\xi_{,\phi}>0$) at the horizon crossing. Substituting the quantities from (\ref{eq: potVal}), (\ref{eq: derV}), and (\ref{eq: derXi}) into (\ref{eq: sqrtbeta}), we find that $\dot{\phi}<0$ during GB inflation. 

Substituting the quantities given in (\ref{eq: potVal}), (\ref{eq: derV}), and (\ref{eq: derXi}) into Eq.~(\ref{eq: forVeff}), we obtain
\begin{align}\label{eq: derVeff}
    \frac{V_{{\rm eff}, \phi}}{M_{\rm pl}^3}
    =\frac{1}{12\sqrt{2}\pi^2 \sqrt{r} \mathcal{P}_S}\left(1+\frac{n_T}{2}\right)^{2}
    \,.
\end{align}
The slope of the effective potential is positive $V_{{\rm eff},\phi}>0$. This implies that the slow-rolling dynamics of the scalar field can be understood in terms of the effective potential in our scenario, which includes the contributions from the potential $V(\phi)$ and the GB coupling function $\xi(\phi)$. The following section examines models to test the validity of these conditions. 

\subsection{Test models}
\label{subsec:toy_models}
Traditionally, models (potentials and couplings) are first handpicked to look at inflationary predictions.
Then, the models are constrained by observational data. However, for inflation,
fine-tuning is arguably required to realize a blue-tilted tensor power spectrum,
since models are able to achieve so at varying degrees.
Alternatively, in this section, we will construct GB inflation models satisfying PTA (Table~\ref{tab:1}) and CMB constraints \cite{Planck:2018jri}, i.e., models with a blue-tilted the tensor power spectrum and a red-tilted scalar power spectrum.
Consequently, we examine two different inflation models: natural inflation with a potential \eqref{eq: NatPot} and power-law GB inflation with a potential \eqref{eq: plpot}. For the former, we demonstrate that the blue-tilted tensor spectrum is produced due to a nontrivial GB coupling, whereas the latter yields only a red-tilted tensor power spectrum.
The blue-tilted spectrum arises because the inflaton field climbs up the potential before rolling down, eventually reaching the end of inflation. In contrast, this climb-up situation does not occur in inflation with power-law potential even with a GB term.

\subsubsection{Natural inflation}\label{sec: M1}
Let us begin with a natural inflation potential~\cite{Freese:1990rb, Freese:2014nla, Stein:2021uge}
\begin{align}\label{eq: NatPot}
    V(\phi)=\Lambda^4\left[1+\cos\left(\frac{\phi}{f}\right)\right]\,,
\end{align}
where $f$ is called the decay constant with dimensions of mass. For the scalar field to climb the potential, or the potential slope to be negative, $V_{,\phi}<0$, we must ensure $\sin(\phi/f)>0$. This is satisfied if $0<\phi<f\pi$. 
Using $n_T$ determined by PTA constraints, we can solve (\ref{eq: nT}) for $\xi(\phi)$ using the potential given in (\ref{eq: srparams}) as
\begin{align}
    \xi(\phi) &= \frac{3M_{\rm pl}^4}{8\Lambda^4} \left[ 2n_{T}^\text{obs} \frac{f^2}{M_{\rm pl}^2}\ln \tan\left(\frac{\phi}{2f}\right)+\sec^2\left(\frac{\phi}{2f}\right)\right]\,, 
\end{align}
where $n_{T}^\text{obs}$ is an observationally favored value of $n_T$ (Table~\ref{tab:1}). The number $N$ of $e$-folds is, therefore, obtained from (\ref{eq: efold}) as
\begin{align}\label{eq: toyNa}
    N&=\frac{2}{n_{T}^\text{obs}}\left[\ln \cos\left(\frac{\phi_e}{2f} \right)-\ln \cos\left(\frac{\phi}{2f} \right)\right] \,,
\end{align}
where 
\begin{align}
    \frac{\phi_e}{M_{\rm pl}}&= \frac{2f}{M_{\rm pl}}\text{arccot}\left(\frac{\sqrt{2}}{n_T^\text{obs}}\frac{M_{\rm pl}}{f}\right)\,.
\end{align}
By inverting (\ref{eq: toyNa}), one can express the scalar field in terms of the number $N$ of $e$-folds as 
\begin{align}\label{eq: phiofN}
    \frac{\phi}{M_{\rm pl}}&=\frac{2f}{M_{\rm pl}} \arccos\left(e^{-\frac{n_T^\text{obs}}{2}N} \cos\left(\frac{\phi_e}{2f}\right)\right)\simeq \frac{2f}{M_{\rm pl}} \arccos\left(e^{-\frac{n_T^\text{obs}}{2}N} \right)\,.
\end{align}
It is straightforward to derive the observable quantities, including $n_S$, $n_T$, and $r$. From (\ref{eq: nSm1}), (\ref{eq: nT}) and (\ref{eq: r}), we obtain
\begin{align}\label{eq: toyM1nsntr}
    \left\{n_S-1, n_T, r\right\} &=\left\{-\frac{2n_T^{\text{obs}}}{2+n_T^{\text{obs}}}\cot^2\left(\frac{\phi}{2f}\right)\,, n_T^\text{obs}\,, 8{n_T^{\text{obs}}}^2\frac{f^2}{M_{\rm pl}^2}\cot^2\left(\frac{\phi}{2f}\right)\right\}
    \,,
\end{align}
where $n_T^\text{obs}$ is the best fit value in Table~\ref{tab:1} and $\phi/M_{\rm pl}$ is given in (\ref{eq: toyM1nsntr}). We can solve (\ref{eq: toyM1nsntr}) for $f/M_{\rm pl}$ and find 
\begin{align}\label{eq: fofMp}
    \frac{f^2}{M_{\rm pl}^2}=\frac{r^\text{obs}}{4n_T^\text{obs}\left( 2+n_T^\text{obs}\right)\left(1-n_S^\text{obs}\right)}\,,
\end{align}
where $n_S^\text{obs}$ is the observationally favored value of the spectral tilt $n_S$ of the scalar perturbations and its value favored by the CMB data is $n_S^\text{obs}=0.9649\pm 0.0042$~\cite{Planck:2018vyg}. 

Substituting the $n_S^\text{obs}$, together with best-fit values of $n_T^\text{obs}$ and $r^\text{obs}$ from Table~\ref{tab:1}, into (\ref{eq: fofMp}), we obtain $f\simeq 0.1556\times 10^{-4} M_{\rm pl}$ for NANOGrav 15-years data and $f\simeq 0.2015  M_{\rm pl}$ for IPTA-DR2 data, respectively, and both values are sub-Planckian. Moreover, we find that the relevant modes (that left the horizon during inflation and gave rise to an enhancement at the PTA scale) must have left the horizon about $\mathcal{O}(1)$ $e$-folds before the end of inflation, i.e.,
$N\simeq 1.63$ for NANOGrav 15-years and $N\simeq2.67 $ for IPTA-DR2 data, respectively. 
At the horizon crossing, we can also estimate $\Lambda$ by using (\ref{eq: potVal}) and obtain $\Lambda\simeq 0.7534\times 10^{-4}M_{\rm pl}$
for NANOGrav 15-years data and $\Lambda\simeq 0.7261\times 10^{-2}M_{\rm pl}$
for IPTA-DR2 data, respectively.  

\subsubsection{Inflation with power-law potential}\label{sec: M2}
Let us now consider inflation with the power-law potential of the form~\cite{1988eur..book..541L}
\begin{align}\label{eq: plpot}
    V(\phi) = V_0\phi^n\,,
\end{align}
where we assume $V_0>0$ and $\phi>0$. To yield the condition that $V_{,\phi}<0$, we find that $n V_0 \phi^{n-1}<0$ should be satisfied, which is true only if $n<0$. Thus, from now on, we will assume $n<0$. To satisfy PTA constraints on $n_T$, the GB coupling function is expected to take the following form 
\begin{align}
    \xi(\phi)&= \frac{3 M_{\rm pl}^4}{4V_0 \phi^n}\left(1+\frac{n_{T}^\text{obs}}{n(n-2)}\frac{\phi^2}{M_{\rm pl}^2} \right)\,,
\end{align}
where $n_{T}^\text{obs}$ is an observationally favored value of $n_T$ (Table~\ref{tab:1}). The number $N$ of $e$-folds is determined from (\ref{eq: efold}) as
\begin{align}\label{eq: plNfoldM1}
    N&=\frac{n}{n_T^\text{obs}} \ln \frac{\phi_e}{\phi}\,,
\end{align}
where $\phi_e$ is obtained by requiring $|\beta(\phi_e)|\equiv 1$ at the end of inflation as 
\begin{align}\label{eq: plphiendM1}
    \frac{\phi_e}{M_{\rm pl}}&=-\frac{\sqrt{2}n}{n_T^\text{obs}}\,,
\end{align}
where noted that $n<0$ is assumed.
Substituting (\ref{eq: plphiendM1}) into (\ref{eq: plNfoldM1}) and inverting the expression for $\phi$, we obtain
\begin{align}
    \frac{\phi}{M_{\rm pl}}&=\frac{\phi_e}{M_{\rm pl}} e^{-\frac{n_T^\text{obs}}{n}N}\,.
\end{align}
The observable quantities of $n_S$, $n_T$, and $r$ for the potential given in (\ref{eq: plpot}) are obtained from (\ref{eq: nSm1}), (\ref{eq: nT}) and (\ref{eq: r}) as
\begin{align}\label{eq: obsM1}
    \left\{n_S-1, n_T, r \right\} &= \left\{ -\frac{2(2-n)n_T^\text{obs}}{n\left(2+n_T^\text{obs}\right)}\,,n_T^\text{obs}\,, \frac{8 n_T^\text{obs}}{n^2}\frac{\phi^2}{M_{\rm pl}^2} \right\}\,,
\end{align}
Since $n_S$ is constrained by CMB data to be $n_S<1$, we find from (\ref{eq: obsM1}) that the power $n$ of the potential in (\ref{eq: plpot}) as 
\begin{align}\label{eq: nval}
    n&=\frac{4n_T^\text{obs}}{\left(2+n_T^\text{obs}\right)\left(1-n_S\right)+2n_T^\text{obs}}\,.
\end{align}
When $n_T^\text{obs}>0$ and $n_S<1$, one can notice that $n>0$. For instance, by substituting the best-fit value of $n_T^\text{obs}$ from Table~\ref{tab:1} and the CMB value of $n_S\simeq0.9649$ into (\ref{eq: nval}), we obtain $n\simeq1.9334$ for NANOGrav 15-years data and $n\simeq 1.9078$ for IPTA-DR2 data, respectively. These positive values of $n$ are in conflict with the negative values of $n$ needed to satisfy the necessary condition to realize a blue-tilted tensor power spectrum, i.e., $V_{,\phi}<0$. Thus, by starting with PTA constraints, we are able to motivate that the blue-tilted tensor spectrum cannot be realized for inflation with a power-law potential, with or without a GB term.

The exception is \cite{Yin:2024ccm} where the GB coupling function was fixed to be of the form of inverse power of the potential. Then, the condition $n<0$ was determined by PTA data, supporting a blue-tilted GW spectrum.

\section{Conclusion} \label{sec: 5}
We have explored the implications of pulsar timing array observations on inflationary models incorporating a Gauss-Bonnet (GB) term. A key result is that GB inflation naturally supports a blue-tilted tensor power spectrum, consistent with current PTA data---a feature difficult to achieve in standard single-field slow-roll inflation due to the consistency relation, $r = -8 n_{T}$.

In GB inflation, the stochastic gravitational wave background depends on the amplitude of a tensor power spectrum, spectral tilt, reheating temperature, and transfer functions---physical quantities that reflect inflationary GWs production and propagation. We derived general conditions for the inflaton potential and GB coupling function, showing that to generate a blue-tilted spectrum, the inflaton must first climb up the potential before rolling down, with a negative potential slope and positive GB coupling slope at horizon crossing. The GB term ensures that slow-roll dynamics remains viable under these conditions.

Our analysis, supported by PTA constraints (Figures~\ref{fig:corner_gb_inflation}, \ref{fig:violins_gb_inflation}, and Table~\ref{tab:1}), highlights key differences from standard inflationary models. We draw the following conclusions. Current PTA data is weakly sensitive to GB-specific parameters because the dominant contributions arise from modes entering the horizon before reheating ends, while PTA data in the relevant ${\cal O}(10^2)$ nanohertz range remains noise-dominated. Reheating temperatures and the tensor-to-scalar ratio in GB inflation are approximately two orders of magnitude larger than the standard case. The tensor spectral index in GB inflation is slightly smaller than in standard inflation, aligning with PTA constraints.

To validate our findings, we examined two concrete GB inflation models (Sections~\ref{sec: M1} and \ref{sec: M2}), demonstrating that the derived conditions hold for a cosine potential but not for a power-law potential. In both cases, the predicted scalar spectral index and tensor-to-scalar ratio remain within observational bounds.

Finally, while PTA data provides valuable insights, fully constraining or ruling out GB inflation will require complementary data across different GW frequency bands, from millihertz (e.g., LISA) to hectohertz (e.g., LIGO), alongside cosmological observations. The framework established here offers a pathway to refining the viable parameter space of GB inflation models in future data analyses. 

{
Gauss-Bonnet inflation can be distinguished from other early universe models that predict a blue-tilted gravitational wave spectrum through several additional observational and theoretical features. One such feature is the presence of distinctive non-Gaussianity patterns, including possible enhancements and unique bispectrum shapes. While non-Gaussianities are often suppressed in other blue-tilted scenarios, their exact signatures remain model-dependent. Another notable characteristic is the possibility of parity-violating gravitational waves, which can arise when a gravitational Chern-Simons term is incorporated alongside the Gauss-Bonnet term in action. Furthermore, Gauss-Bonnet inflation benefits from a natural embedding in string theory, offering a well-motivated ultraviolet completion, unlike many alternative models. Comparing predictions across multiple observables---such as CMB polarization, the stochastic gravitational wave background (e.g., LISA, DECIGO), and reheating dynamics---provides a promising path to distinguish Gauss-Bonnet inflation from competing frameworks like bouncing cosmologies or string gas models. Identifying and constraining these unique signatures remains a key goal in theoretical and observational cosmology
}

\acknowledgments
The authors thank Pritha Bari and Guillem Dom\'{e}nech for comments on a preliminary draft, and to Lu Yin and Jiho Yang for important discussion during this work's initial stages. We thank participants of the external program APCTP-GW2025 [or APCTP2025-E05] held at Academia Sinica, Taipei, Taiwan for fruitful discussions. RCB is supported by an appointment to the JRG Program at the APCTP through the Science and Technology Promotion Fund and Lottery Fund of the Korean Government and was also supported by the Korean Local Governments in Gyeongsangbuk-do Province and Pohang City.
RCB acknowledges the hospitality of the Institute of Physics, Academia Sinica that enabled the completion of this work. SK and GT acknowledge support by the Basic Science Research Program through the National Research Foundation of Korea (NRF), funded by the Ministry of Education, (grant numbers) (NRF-2022R1I1A1A01053784), (NRF-2021R1A2C1005748).


\providecommand{\href}[2]{#2}\begingroup\raggedright\endgroup

\end{document}